\documentclass{revtex4}
\usepackage{graphicx}
\usepackage{dcolumn}
\usepackage{amsmath}
\usepackage{amssymb}
\usepackage{amsfonts}
\usepackage{bm}
\usepackage{color}
\newcommand{\be}{\begin{equation}}
\newcommand{\ee}{\end{equation}}
\newcommand{\ba}{\begin{eqnarray}}
\newcommand{\ea}{\end{eqnarray}}

\begin{document}

\title{A didactic approach to the Machine Learning application to weather
forecast}

\author{Marcello Raffaele$^{\dag}$}
\author{Maria Teresa Caccamo$^{\dag}$}
\author{Giuseppe Castorina$^{\dag}$} 
\author{Gianmarco Muna\`o$^{\dag}$}
\author{Salvatore Magaz\`u$^{\dag}$}
\thanks{Corresponding author, email: {\tt smagazu@unime.it}}

\affiliation{
$^{\dag}$Dipartimento di Scienze Matematiche e Informatiche, Scienze Fisiche
e Scienze della Terra,
Universit\`a degli Studi di Messina,
Viale F.~Stagno d'Alcontres 31, 98158 Messina, Italy.} 

%

\maketitle

\section*{Abstract}
We propose a didactic approach to use the Machine Learning protocol in order
to perform weather forecast. This study is motivated by the possibility to 
apply this method to predict weather conditions in proximity of the Etna
and Stromboli volcanic areas, located in Sicily (south Italy). 
Here the complex orography may significantly influence the
weather conditions due to Stau and Foehn effects, with
possible impact on the air traffic of the nearby Catania and Reggio Calabria 
airports. 
We first introduce a simple thermodynamic approach, suited to provide 
information on temperature and pressure when the Stau and Foehn effect takes
place. In order to gain information to the rainfall accumulation, the Machine
Learning approach is presented: according to this protocol, the model is able
to ``learn'' from a set of input data which are the meteorological conditions
(in our case dry, light rain, moderate rain and heavy rain) associated to
the rainfall, measured in mm. We observe that, since in the input dataset 
provided by the Salina weather station the dry condition was the most common,
the algorithm is very accurate in predicting it. Further improvements can
be obtained by increasing the number of considered weather stations and time
interval.

\section{Introduction}
\begin{figure}[t!]
\begin{center}
\includegraphics[width=11.0cm,angle=0]{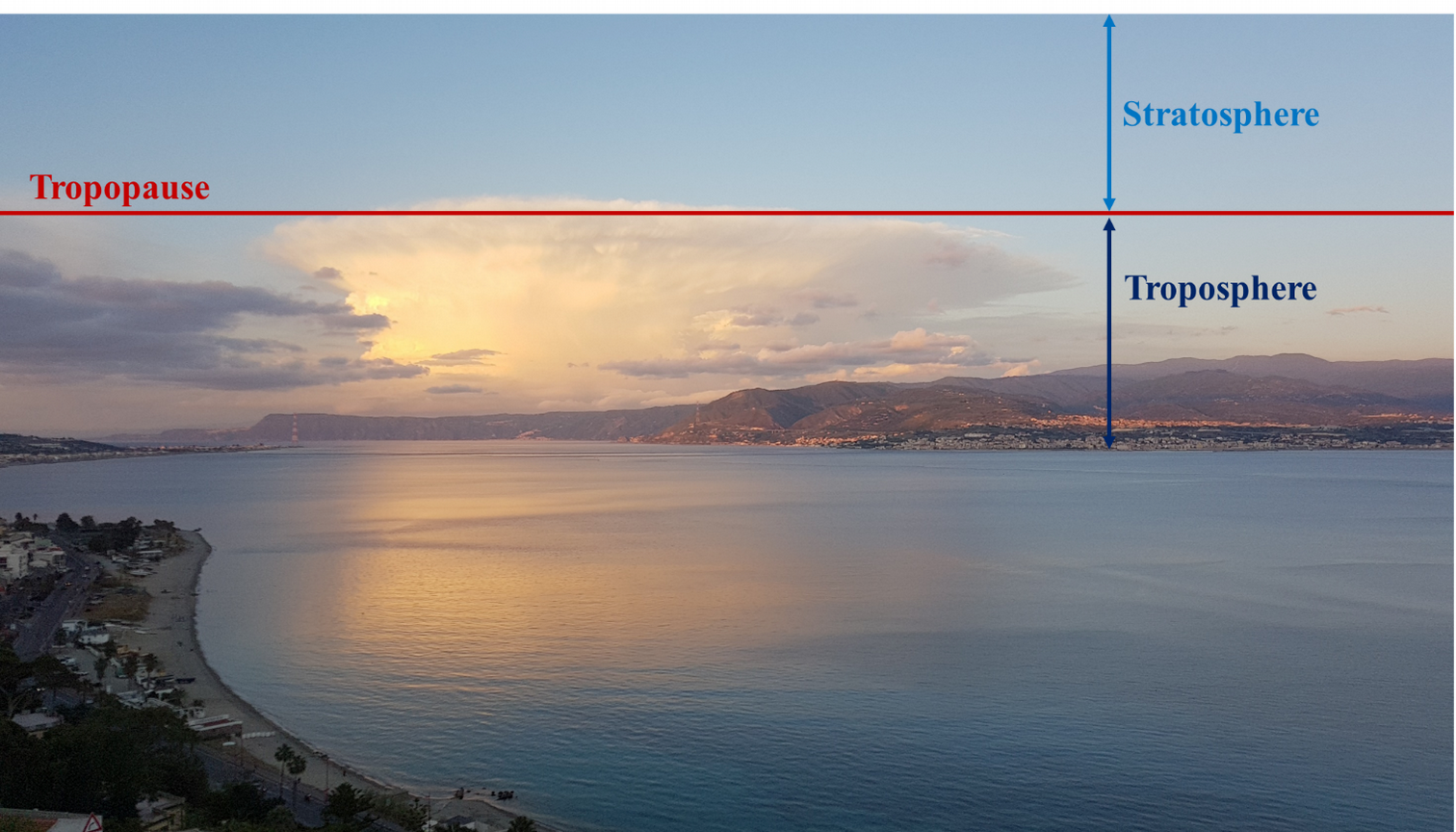}
\caption{Visual representation of the troposphere and stratosphere, along
with the separation between them, called tropopause.}\label{fig:ML0}
\end{center}
\end{figure}

The Earth's atmosphere is a part of a very complex system, known as Earth 
System Sciences. In addition to the atmosphere, it also includes the 
hydrosphere (the water envelope formed by seas, rivers, lakes and underground 
waters), the cryosphere (the part of the Earth's surface that is covered from 
the ice), the biosphere (the set of areas of the Earth in which the conditions 
necessary for animal and plant life exist) and the lithosphere 
(the outermost 
part of the Earth, formed by two layers, the crust and the 
mantle)~\cite{Wallace:77}.
Generally, the atmosphere is divided into different layers on the basis of the 
vertical profile of the temperature (temperature gradient) or on the basis of 
particular physical or chemical phenomena characterizing it. 
The temperature gradient allows one to identify the various atmospheric 
layers on the basis of the different characteristics between 
them~\cite{Orlanki:75}. 
It is possible to identify two macro layers of the atmosphere: in the 
heterosphere (altitude $\geq$ 100 km) the average molecular free path 
is greater 
than one meter. Under these conditions the concentration of the heavier 
elements decreases with the altitude, compared to the lighter ones and there 
is no dependence with the vertical profile of the temperature. In the 
omosphere (altitude $\leq$ 100 km) the concentration of the main constituents 
tends to be uniform and independent of altitude due to the turbulent mixing; 
there is dependence on the vertical temperature profile.
In the omosphere it is possible to identify the following layers of the 
atmosphere: troposphere, stratosphere, and mesosphere. A pictorial view
of troposphere and stratosphere is reported in Fig.~\ref{fig:ML0}, where
the separation between them (tropopause) is also visible. 
Among the different atmospheric layers, 
the troposphere plays a fundamental role since it contains 
the eighty percent of the total mass of the atmosphere; it also contains 
almost all the water vapor and, therefore, is the location where 
meteorological phenomena develop~\cite{Fletcher:62}.
Usually, in the troposphere there is a constant decrease in temperature with
the altitude: this is due both to the possibility for an atmospheric layer to 
expand adiabatically (less pressure and therefore cooling) and, expecially, to
the fact that the main source of heat is provided by the soil. 
The decreasing trend of the temperature with the altitude in the troposphere 
is the basis of the condensation phenomenon of the water vapor due to the 
orographic forcing. In this context, when the motion of mass of air is hampered 
by the presence of an orographic obstacle, the mass it is forced to rise. 
In this case, as the altitude increases, the temperature decreases. This 
cooling causes the condensation of water vapor thus favoring the genesis of 
extensive cloud systems with precipitation on the windward side (Stau effect). 
Then, when the air mass reaches the leeward slope, it starts a downward motion, 
with a corresponding increse of the pressure and therefore heating by 
adiabatic compression (Foehn effect). These effects occur especially in areas 
with a complex orography such as that which characterizes the Sicily region. 
The most important examples in this context are constituted by 
Etna and Stromboli volcanic areas. 
In particular, with its height of 3300 meters and its proximity to 
the Peloritani and Nebrodi mountains, Etna can favor the 
cooling and subsequent heating of the air masses, thus acting as a trigger for 
the genesis of even extreme weather events. Indeed, in the Ionian area of 
Sicily, it is not unusual that rainfall accumulations on the ground close to 
300 mm in 2-3 hours occur~\cite{MTCaccamo:17}. 
In principle, a proper knowledge of the nucleation 
processes~\cite{Restuccia:18} and of cloud 
microphysics~\cite{Castorina:19} could be significantly helpful, 
but, due to the complex weather
conditions, this is not a straightforward task.
In this context, the need to develop novel 
approaches suited to predict or reproduce such weather events clearly 
emerges. \\
The aim of the present work is twofold: from the one hand, we plan to provide
a didactic explanation of the complex weather phenomena usually happening
in region characterized by a complex orography, such as the Etna 
and Stromboli volcanic
areas. From the other hand, we also develop a novel strategy to perform
weather forecast avoiding the implementation of complex mathematical models
and making use of the Machine Learning approach.
The Machine Learning technique uses data to identify useful information 
without needing to know any mathematical formulas or specific codes, 
since the algorithm generates its own logic, based on the data entered in 
input and output~\cite{Alpaydin:20}. 
One of these algorithms is called classification: it can 
insert data into different groups based on common characteristics that it 
is able to identify them independently. Learning can be supervised 
or unsupervised: the first one must know the previous answers to the problem 
(also called training data) and is able to work backwards to understand the 
logic between input and output. Instead, the unsupervised has no known 
answers to the problem used for training and the training set is not labeled 
as in the supervised case. The Machine Learning protocol has been recently
implemented to predict weather forecast uncertainty~\cite{Scher:18},
measure raindrops~\cite{Denby:01}, perform drought forecasting for ungauged 
areas~\cite{Rhee:17} and apply nowcasting methods based on real‐time 
reanalysis data~\cite{Han:17}. \\
In this work, we illustrate how to apply the Machine Learning 
techniques to the weather forecasting. The input signals are get by sensors  
providing data related to the most important physical quantities (for instance 
temperature, pressure, rainfall, etc.) taken from fixed stations 
located in Sicily. Data concerning the output signals
are related to the weather forecast of the days at issue, 
taking into account that the weather condition is affected by the 
input data. The work is organized as follows: in the next section
a thermodynamic approach to describe the Stau and Foehn effects is presented;
the Machine Learning technique is described in Section 3 and the results are 
presented in Section 4. Conclusions follow in the last section.

\section{Weather conditions and complex orography: the cases of the Stau and 
Foehn effects}
A forced lifting occurs when a moving mass of air is forced to rise in front 
of an orographic obstacle (forced orographic ancestry). Lifting speeds are 
in the range of [0.5 - 1] m/s, with a decrease in temperature in the unit 
of time greater than that observed in large baric centers.
The cooling, in general, causes the condensation of water vapor with 
extensive cloud formations and rainfall on the windward side. If the air is 
initially unsaturated, its upward movement takes place along a dry 
adiabatic, with a cooling rate of 1 $^\circ$C every 100 meters, up to the level 
where this cooling does not produce condensation: indeed the heat released 
by the condensation attenuates the cooling of the rising air.
The lifting continues according to the saturated adiabatic with a thermal 
gradient that depends on the initial values of temperature, specific humidity 
and ascent rate. A realistic value for this thermal gradient is approximately 
0.5 - 0.6 $^\circ$C per 100 meters.
The upward movement of the air on the windward side of a mountain range 
(for example the Etna and Stromboli volcanic areas), 
with the formation of clouds and 
rainfall is called the Stau effect. In this phase, the abundant rainfall dry 
the rising air mass. When the latter crosses the leeward slope, in its 
downward motion, it undergoes an adiabatic compression with a corresponding 
heating of 1 $^\circ$C every 100 meters.
This heat gain, not used to re-evaporate the clouds formed in the ascent phase 
(now dry air masses) is absorbed entirely by the air mass, which therefore 
reaches ground in a warmest and driest condition than it was originally.
This is know as Foehn effect.
The adiabatic expansions and compressions are well known examples 
of thermodynamic processes and have been
recently investigated also by means of the R{\"u}chardt's 
experiment~\cite{MTCaccamo:19} and the frequency analysis 
procedure~\cite{Castorina:18}.
The Stau and Foehn effects can be thermodynamically described by 
making use of the
first thermodynamic law, which can be written as:
\begin{equation}\label{eq:du}
\Delta U = Q -L \,,
\end{equation}
where $\Delta U$ indicates the variation of the internal energy of a 
thermodynamic system and $Q$ and $L$ are the amount of heat adsorbed by the
system and the work performed by the system on the surrounding environment.
Since, in the case of the Stau and Foehn effects, the thermodynamic 
process takes place
without exchanging heat with the environment ({\it i.e.} adiabatically),
$Q=0$. Therefore, Eq.~\ref{eq:du} amounts to:
\begin{equation}\label{eq:du1}
\Delta U = -L \,.
\end{equation}
Since both the effects can be studied through an adiabatic process,
for the sake of simplicity here we discuss in detail the Foehn effect only.
According to the Foehn effect, the altitude decreases during the process, and
hence the pressure increases, this leading to an adiabatic compression. 
Assuming the approximation of the ideal gases, the infinitesimal variation
of internal energy can be written as follows:
\begin{equation}\label{eq:duvar}
dU= n C_V dT \,,
\end{equation}
where $n$ is the number of moles, $C_V$ the specific heat at constant 
volume and $dT$ the infinitesimal variation of the temperature.
Since the work performed by the systems corresponds to its pressure multiplied
by the volume change, combining Eqs.~\ref{eq:du1} and ~\ref{eq:duvar} 
we obtain:
\begin{equation}\label{eq:work}
pdV=dL=-dU=- n C_V dT \,.
\end{equation}
The equation of perfect gases can be written in a differential form as:
\begin{equation}\label{eq:diff}
d(pV)=d(nRT) \rightarrow pdV + Vdp = nRdT \,,
\end{equation}
where $R$ is the universal gas constant. By using Eq.~\ref{eq:work}, 
Eq.~\ref{eq:diff} may be rewritten as:
\begin{equation}\label{eq:vdp}
Vdp = nRdT + n C_V dT \,.
\end{equation}
Eq.~\ref{eq:vdp} my be expressed in terms of the specific heat at constant
pressure, $C_P$, which is defined as:
\begin{equation}\label{eq:cp}
C_P = R + C_V \,.
\end{equation}
Therefore, Eq.~\ref{eq:vdp} becomes:
\begin{equation}\label{eq:vdp1}
Vdp = n C_P dT \,.
\end{equation}
By dividing Eq.~\ref{eq:vdp1} by Eq.~\ref{eq:work} we obtain:
\begin{equation}\label{eq:gamma}
\frac{Vdp}{pdV} = -\frac{n C_P dT}{n C_V dT} = -\frac{C_P}{C_V} = 
-\gamma \,,
\end{equation}
where $\gamma$ is defined as the ratio $C_P/C_V$.
By rearranging Eq.~\ref{eq:gamma}, a useful relation between pressure and
volume can be found:
\begin{equation}\label{eq:dpdV}
\frac{dp}{p} = -\gamma \frac{dV}{V} \,,  
\end{equation}
which, after integration, provides the result:
\begin{equation}\label{eq:log}
{\rm ln} \frac{p_f}{p_i} = -\gamma {\rm ln} \frac{V_f}{V_i} \,,  
\end{equation}
where the indexes $f$ and $i$ label the final and initial state, respectively.
Eq.~\ref{eq:log} can be rearranged in turn as:
\begin{equation}\label{eq:exp}
p_i V_i^{\gamma} = p_f V_f^{\gamma} \,,
\end{equation}
which is known as Poisson equation; the latter, finally, leads to the 
expression:
\begin{equation}\label{eq:const}
p V^{\gamma} = constant \,,
\end{equation}
which provides the relation between pressure and volume in the course of 
the adiabatic transformation which takes places during the Foehn effect. 
However, Eq.~\ref{eq:const} does not provide any information on the 
temperature. The latter can be obtained by combining Eq.~\ref{eq:exp}
with the equation of the perfect gases:
\begin{equation}\label{eq:gas}
pV = nRT \,.
\end{equation}
From Eq.~\ref{eq:exp} and Eq.~\ref{eq:gas} we obtain:
\begin{equation}\label{eq:new}
p_i \bigg[\frac{nRT_i}{p_i}\bigg]^{\gamma} = p_f \bigg[\frac{nRT_f}{p_f}\bigg]
^{\gamma} \,,
\end{equation}
which can be rewritten as:
\begin{equation}\label{eq:pt}
p_i^{1-\gamma}T_i^{\gamma}=p_f^{1-\gamma}T_f^{\gamma} \,.
\end{equation}
A proper rearrangement of Eq.~\ref{eq:pt} provides:
\begin{equation}\label{eq:pt1}
T_i p_i^{\frac{1-\gamma}{\gamma}}=T_f p_f^{\frac{1-\gamma}{\gamma}} \,.
\end{equation}
For the sake of simplicity, we define a new constant 
$\alpha\equiv \frac{1-\gamma}{\gamma}$; therefore, Eq.~\ref{eq:pt1} becomes:
\begin{equation}\label{eq:alpha}
T_i p_i^{\alpha}=T_f p_f^{\alpha} \,.
\end{equation}
By taking the logarithm of both the members of Eq.~\ref{eq:alpha} we obtain:
\begin{equation}\label{eq:lnalpha}
{\rm ln}(T_i) +\alpha {\rm ln}(p_i)={\rm ln}(T_f) +\alpha {\rm ln}(p_f) \,,
\end{equation}
which can be rearranged into:
\begin{equation}\label{eq:lnt1}
{\rm ln}\bigg(\frac{T_i}{T_f}\bigg) =\alpha {\rm ln}\bigg(\frac{p_f}{p_i}\bigg) \,.
\end{equation}
Now we can rewrite $\alpha$ as follows:
\begin{equation}\label{eq:newalfa}
\alpha\equiv \frac{1-\gamma}{\gamma}=\frac{1-C_P/C_V}{C_P/C_V} 
=\frac{C_V-C_P}{C_P} \,.
\end{equation}
Since $C_V=C_P-R$, Eq.~\ref{eq:newalfa} finally becomes:
\begin{equation}\label{eq:alfalast}
\alpha=\frac{C_P-R-C_P}{C_P}=-\frac{R}{C_P} \,.
\end{equation}
Combining Eq.~\ref{eq:alfalast} with Eq.~\ref{eq:lnt1} we obtain:
\begin{equation}\label{eq:lncp}
{\rm ln}\bigg(\frac{T_i}{T_f}\bigg) =-\frac{R}{C_P} 
{\rm ln}\bigg(\frac{p_f}{p_i}\bigg) \,,
\end{equation}
which, after rearranging the second member, finally turns into:
\begin{equation}\label{eq:last}
{\rm ln}\bigg(\frac{T_i}{T_f}\bigg) =\frac{R}{C_P} 
{\rm ln}\bigg(\frac{p_i}{p_f}\bigg) \,.
\end{equation}
Eq.~\ref{eq:last} provides a clear relationship between the values of
temperature and pressure at the beginning and at the end of the adiabatic
transformation and therefore is particularly useful in order to gain
knowledge on these thermodynamic variables during the process.
In particular, upon setting $p_i= 1000$ hPa, corresponding to the atmospheric
pressure at the sea level, it is possible to introduce the concept of
potential temperature $\Phi$:
\begin{equation}\label{eq:phi}
\Phi =T_f\frac{p_0}{p_f}{\rm exp}\bigg[\frac{R}{C_P}\bigg] \,. 
\end{equation}
The potential temperature indicates the final temperature reached by an 
atmospheric layer during an adiabatic expansion or compression from a given
initial state till to the sea-level pressure.
An important consequence of the definition of potential temperature is that
it does not depend on the altitude $z$, {\it i.e.}
\begin{equation}\label{eq:dphi}
\frac{d\Phi}{dz} = 0 \rightarrow \Phi=constant \,. 
\end{equation}
As a consequence, its value does not change during the whole adiabatic process.
A surface defined by Eq.~\ref{eq:dphi} is known as isentropic surface and 
an atmospheric layer which lies on such a surface will remain on it in the
course of the adiabatic transformation. Further details on the way to obtain
the potential temperature, along with its numerous applications, can be 
found in Ref.~\cite{Molders:14}. \\
It emerges that, under adiabatic conditions, many information on the weather
conditions can be gained by using simple thermodynamic relationships. On 
the other hand, this approach can not provide information on other physical
quantities, such as rainfall accumulations or wind speed, that are of great
importance for a proper determination of the weather conditions. For such 
an aim, in parallel to the thermodynamic approach, we propose a novel
methodology for the weather forecast, based on the Machine Learning protocol,
which is discussed of the next section.

\section{The Machine Learning approach}

\begin{figure}[t!]
\begin{center}
\includegraphics[width=8.0cm,angle=0]{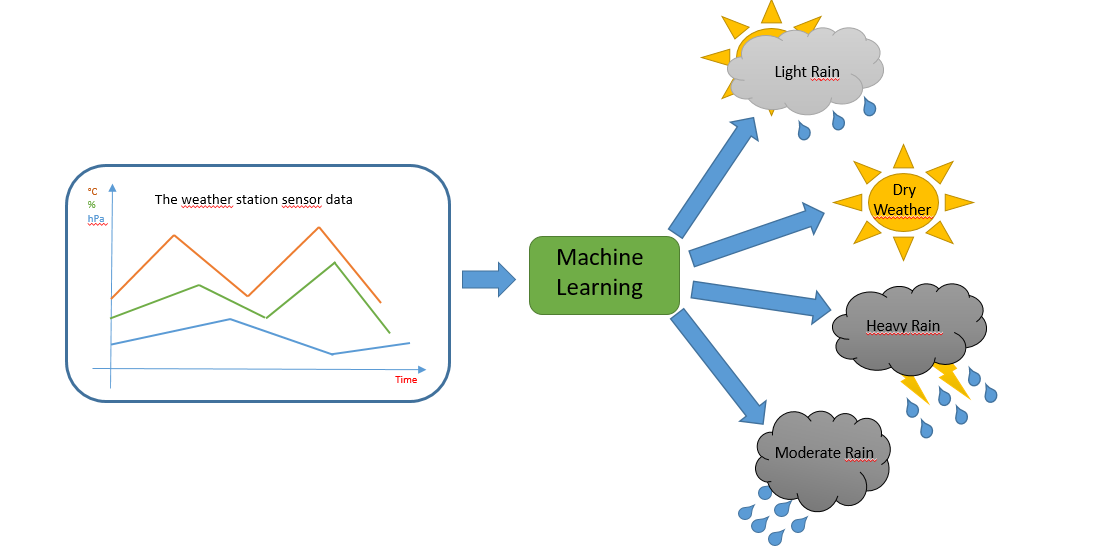}
\caption{Illustration of the four different weather conditions investigated 
in this work by using the Machine Learning approach.}\label{fig:ML1}
\end{center}
\end{figure}

\begin{figure}[t!]
\begin{center}
\includegraphics[width=8.0cm,angle=0]{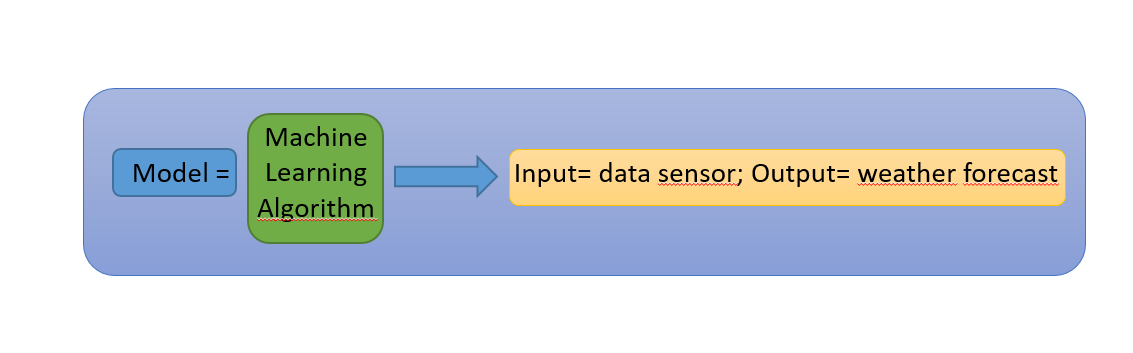}
\caption{Schematic representation of the Machine Learning workflow.
}\label{fig:ML2}
\end{center}
\end{figure}

\begin{figure}[t!]
\begin{center}
\includegraphics[width=8.0cm,angle=0]{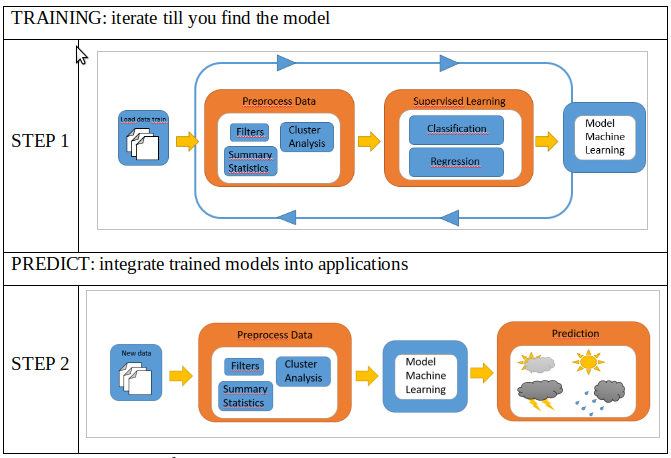}
\caption{Illustration of the two steps of training and prediction.
}\label{fig:ML3}
\end{center}
\end{figure}

\begin{figure}[t!]
\begin{center}
\includegraphics[width=8.0cm,angle=0]{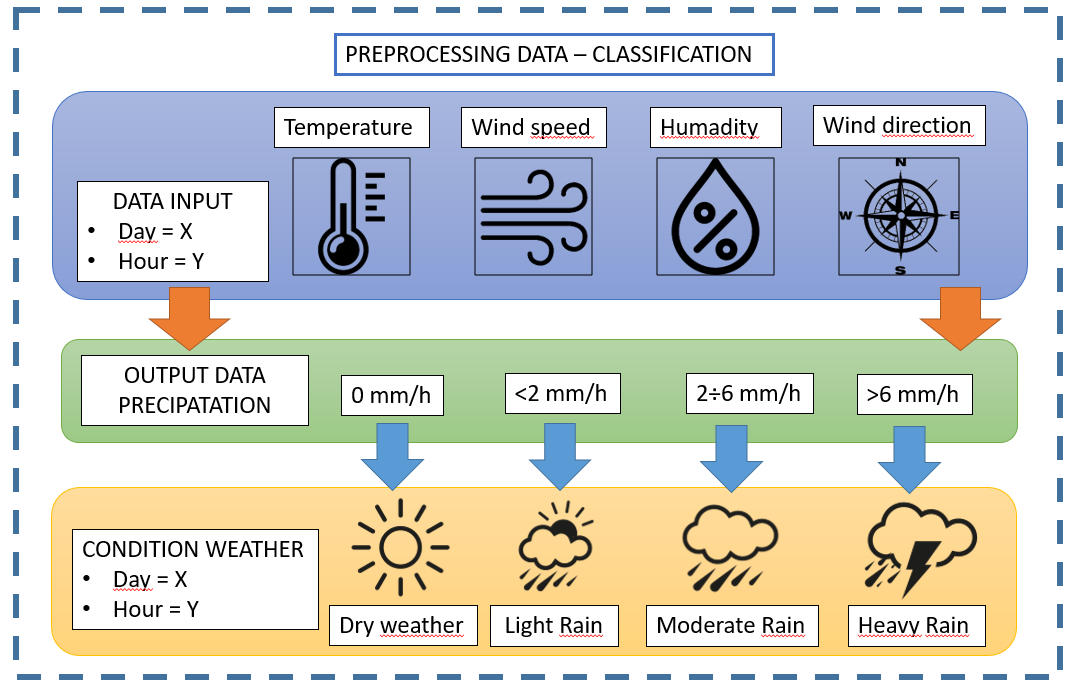}
\caption{Schematic representation of the data organization according to the
Machine Learning protocol.}\label{fig:ML5}
\end{center}
\end{figure}

\begin{figure}[t!]
\begin{center}
\includegraphics[width=8.0cm,angle=0]{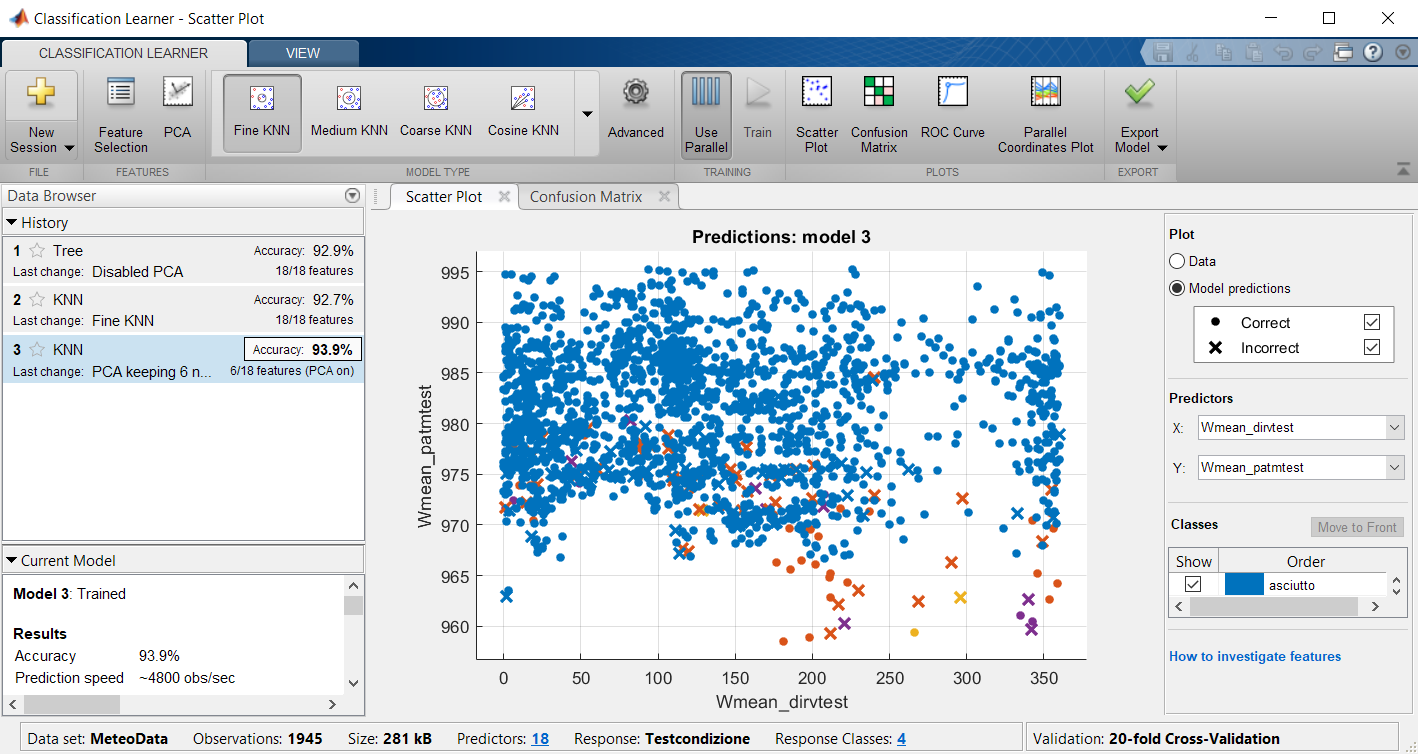}
\caption{Prediction accuracy of the Classification Learner KNN.}\label{fig:ML7}
\end{center}
\end{figure}

The Machine Learning protocol is schematically depicted in
Fig.~\ref{fig:ML1}.
As can be seen from the graphical representation, the approach learns directly 
from the data. Specifically, we provide the input and output data to the
algorithm and the software ``will have to learn'' how to solve the problem; 
this step is defined as a real training. The model so obtained can be used to 
define 
the meteorological activity which will be defined in four different
conditions: \\
1. Dry; \\
2. Light rain; \\	
3. Moderate rain; \\
4. Heavy rain. \\
A block illustration of the model workflow is provided in 
Fig.~\ref{fig:ML2}: a proper definition of the model is not a simple task, 
since data
can come from many different sources, including 
sensors, images or database. Another important aspect concerns the 
data preprocessing, 
which requires specific algorithms for a specific application domain. 
For instance, if the input data are constituted by images, 
we must use algorithms 
that extract the features, whereas if they are constituted by historical 
series, statistical algorithms are needed. A proper choice of the most
accurate algorithm among the large amount of possible protocols can
require long time; usually, the best model is that one which guarantees a 
good balance between speed, accuracy and complexity.
The last step of the procedure requires to iteratively repeat the workflow 
that led us to identify the best model.
We can divide the workflow into two steps, as shown in Fig.~\ref{fig:ML3}. 
The first step is the training, where the data coming from the 
weather stations are first processed by using statistical methods, and then
a classifier is applied on the processed data, in order to build the model. 
The latter is then obtained through several iteration cycles. During the 
second step we use the trained model to perform the prediction, and therefore 
a new dataset is provided for testing the neural network. 
In this context it is worth to point out the importance to have as much data 
as possible. In the present work, the input data for the training phase 
are physical quantities 
determining the weather conditions and are listed below: \\
1. Wind speed [m/s] and direction [$^\circ$]; \\ 
2. Atmospheric pressure [hPA]; \\
3. Maximum and minimum humidity [\%]; \\
4. Maximum and minimum temperature [$^\circ$C]; \\
5. Rainfall [mm]. \\
These data have been recorded by the weather station 
located on the island of Salina 
and have been provided by SIAS (Sicilian Agrometeorological Information 
Service), and collected with an hourly frequency for the days considered. 
\begin{figure}
\begin{center}
\begin{tabular}{ccc}
\includegraphics[width=2.9cm,angle=0]{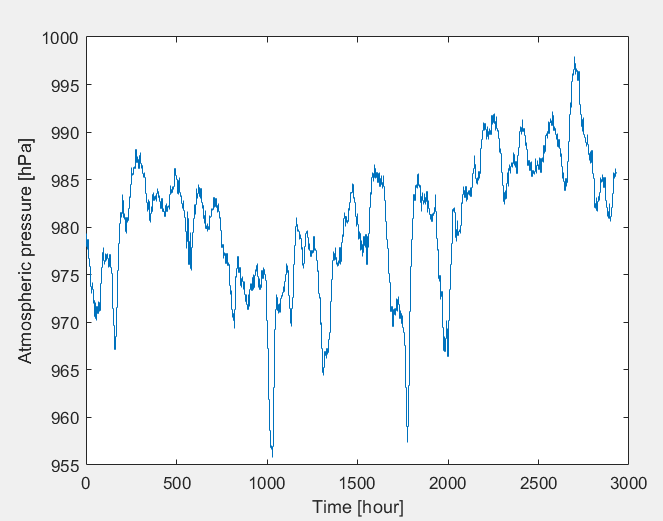}
\includegraphics[width=3.0cm,angle=0]{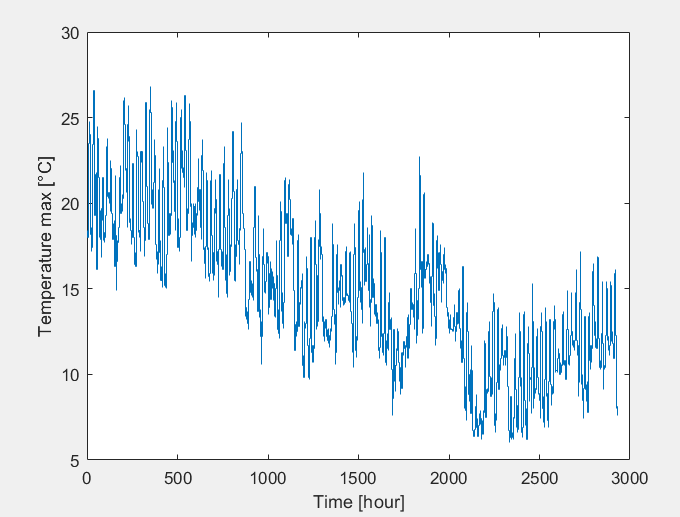} 
\includegraphics[width=2.8cm,angle=0]{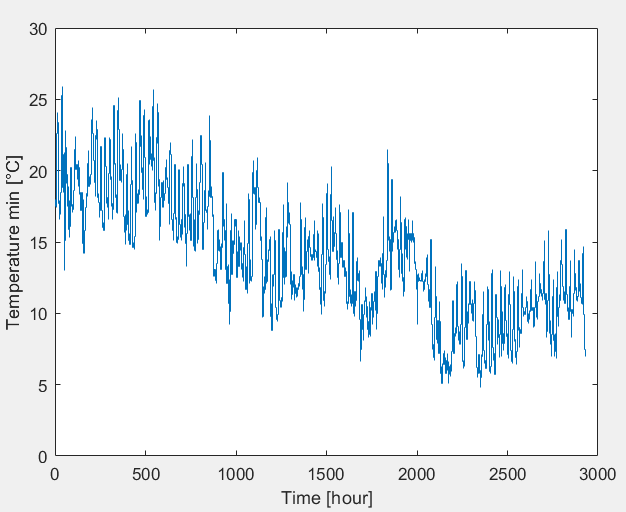} \\
\includegraphics[width=3.0cm,angle=0]{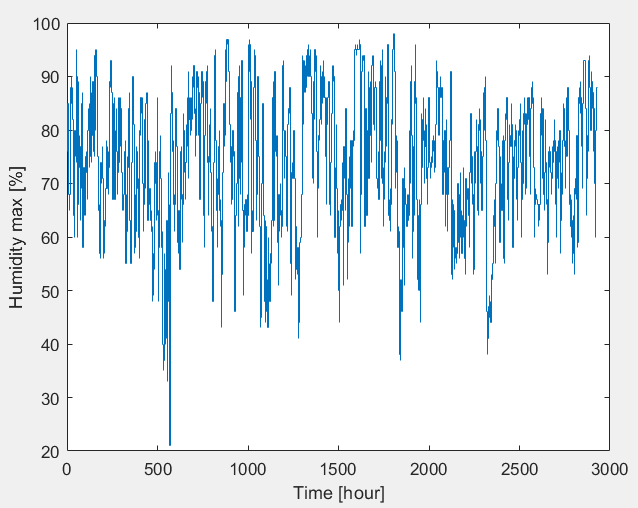} 
\includegraphics[width=3.0cm,angle=0]{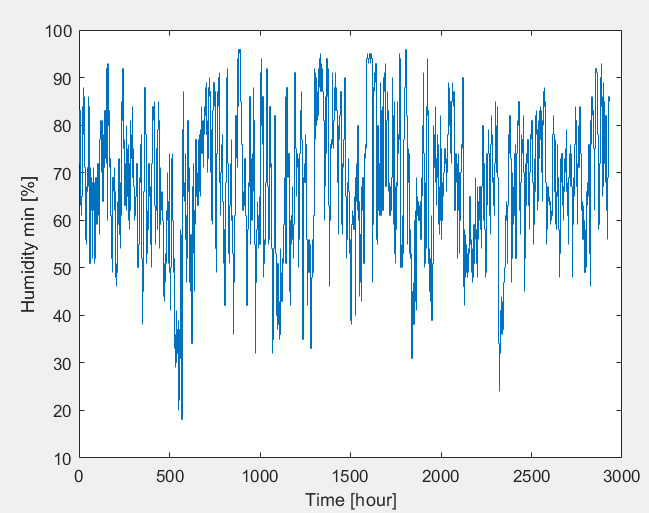} \\
\includegraphics[width=3.0cm,angle=0]{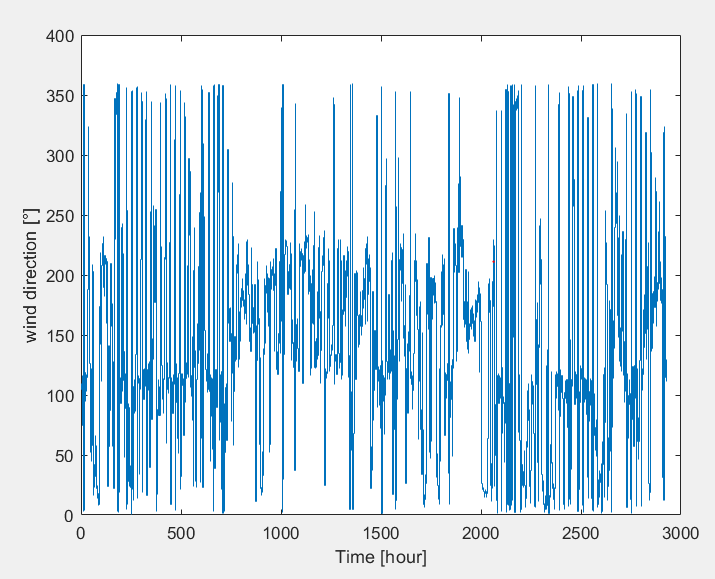}
\includegraphics[width=3.0cm,angle=0]{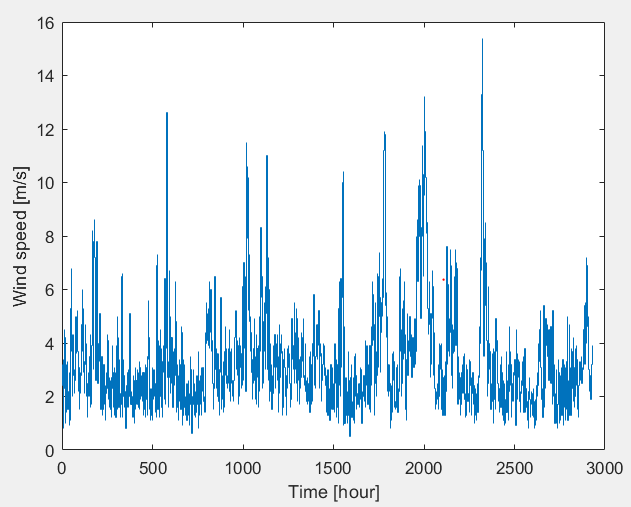}
\end{tabular}
\caption{Input data for the model training obtained from the Salina weather
station.}\label{fig:ML4}
\end{center}
\end{figure}
Such data, plotted in Fig.~\ref{fig:ML4},
cover a time interval going from 
10/01/2019 to 01/31/2020, collecting 2929 total measurements.
The data implemented for the functional test of the Machine Learning algorithm 
are 1945 measurements made from 02/01/2020 to 04/21/2020.
The Machine Learning approach adopted in the present work is defined as a 
classification method and the algorithm is called supervised; in the training 
phase, this type of model needs to know the answers, i.e. the output of a 
given event. The input data of the system are provided by the Salina weather 
station which records every hour of every day certain quantities such as 
temperature, humidity, pressure, speed and wind direction. The output data 
that provide the wanted weather condition on a certain hour of a certain day 
are given by the measured rainfall. In particular, we have set the following
matches~\cite{Giuliacci:03}:\\
1. Dry occurs when there is no rainfall;\\
2. Low rain occurs when rainfall is less than 2 mm/h;\\
3. Moderate rain occurs when rainfall is between 2 and 6 mm/h;\\
4. Heavy rain occurs when rainfall is greater than 6 mm/h.\\
By using these information, the input data are classified through the weather 
condition, so that the system can be trained to recognize which physical 
quantities come into play in the determination of a rainfall. For such an aim, 
an Excel file was organized with all the data imported from the Salina weather 
station. The workflow for organizing such data is reported in 
Fig.~\ref{fig:ML5}.
\begin{figure}[t!]
\begin{center}
\includegraphics[width=8.0cm,angle=0]{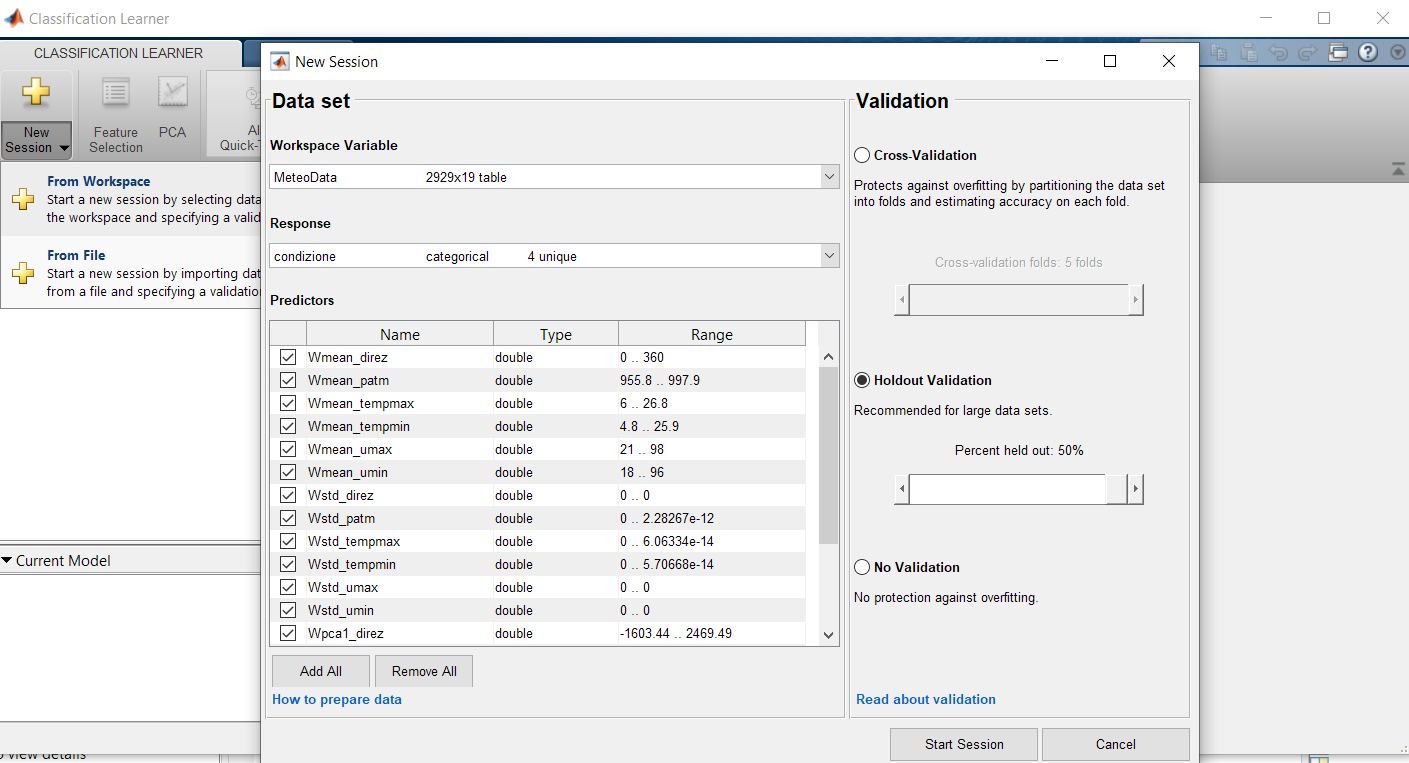}
\caption{Schematic description of the Classification Learner K-Nearest 
Neighbours (KNN).}\label{fig:ML6}
\end{center}
\end{figure}
After the data organization, the next step concerns the data preprocessing: 
for such an aim, the Matlab software has been implemented. In particular, 
after importing the data into the workspace, it is necessary to manipulate 
these data in order to make it easily usable by our Machine Learning 
algorithms, through extraction of features; the extraction has been performed
by means of the 
statistical laws, and in particular by using average value functions, standard 
deviation functions and functions of the analysis of the main components.
After performing this transformation, we need to create a table that includes 
all the data. The next step is the opening of the Classification Learner which 
is a Matlab function whose task is to import the table containing the data 
(see Fig.~\ref{fig:ML6}). In this phase it is possible to choose the method 
for the data validation among two possible 
mechanisms: the first one is called cross-validation and is used in case 
of few available
data, with the software trying to make more efficient use of such data.
The second case is known as holdout validation and allows one to select some 
data for the validation and the remaining amount of data for the training 
phase; in this study, we have applied both the methods with the same frequency.
When the data are read by the classifier, it is possible to choose the 
algorithm among those selectable to start the training phase. In 
Fig.~\ref{fig:ML7} it is possible to note the percentages that refer to the 
prediction accuracy of the algorithm defined as K-Nearest Neighbours 
(KNN)~\cite{Liu:16}.

%
\section{The Machine Learning predictions}

\begin{figure}
\begin{center}
\begin{tabular}{ccc}
\includegraphics[width=2.9cm,angle=0]{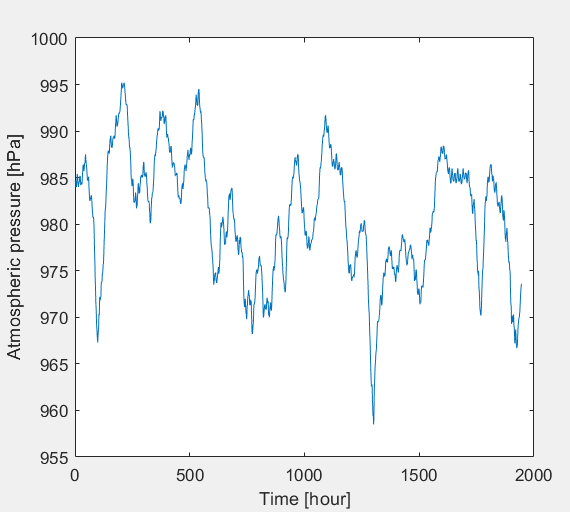}
\includegraphics[width=3.0cm,angle=0]{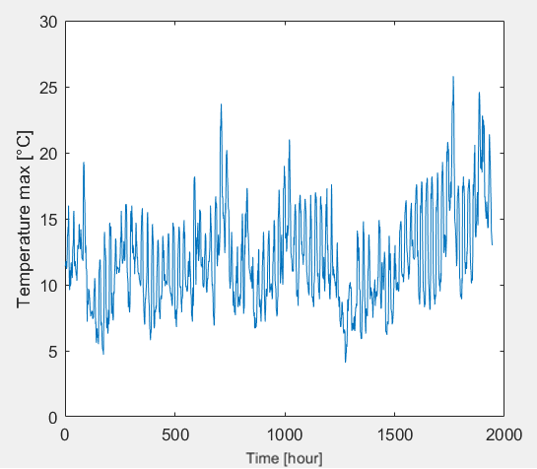} 
\includegraphics[width=2.8cm,angle=0]{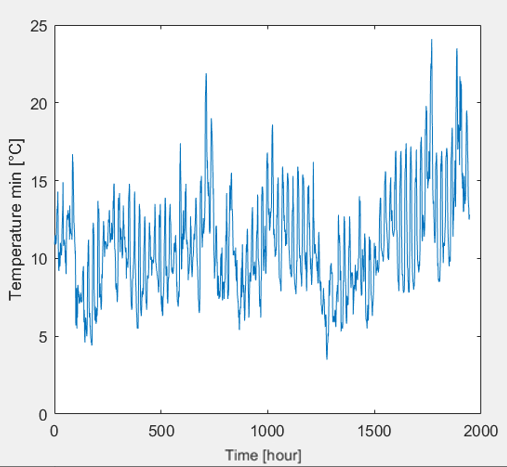} \\
\includegraphics[width=3.0cm,angle=0]{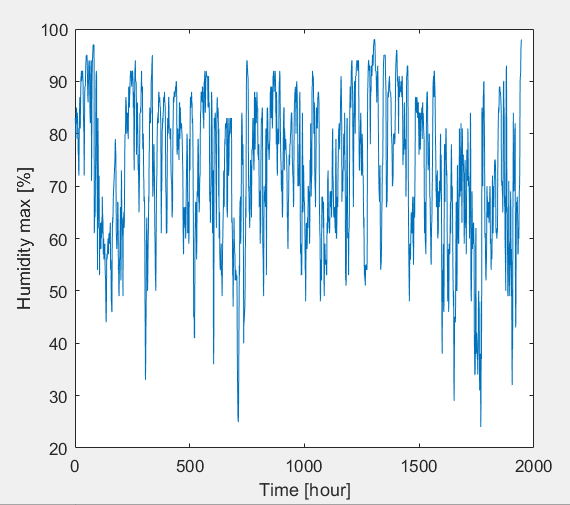} 
\includegraphics[width=3.0cm,angle=0]{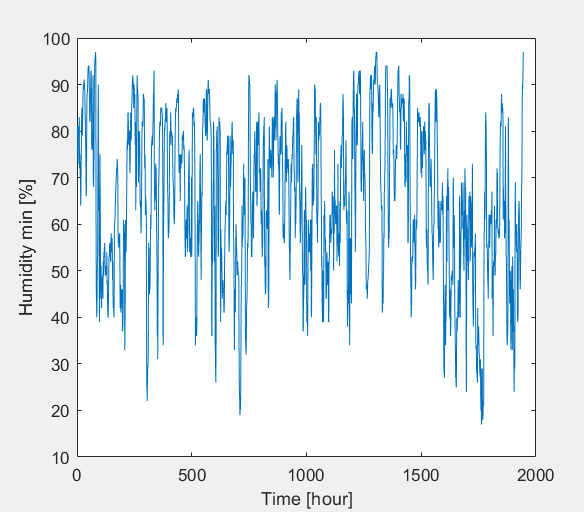} \\
\includegraphics[width=3.0cm,angle=0]{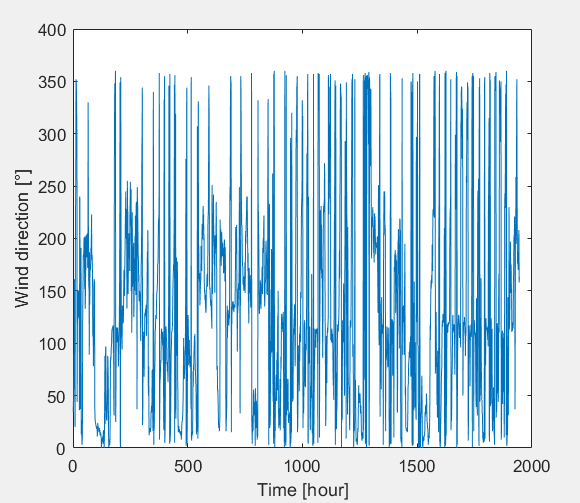}
\includegraphics[width=3.0cm,angle=0]{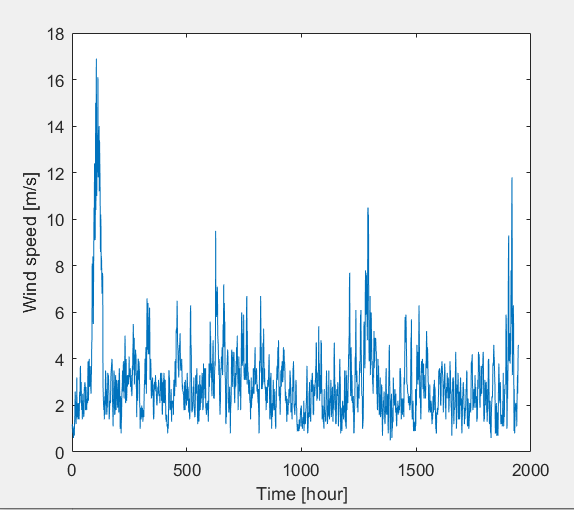}
\end{tabular}
\caption{Input data for the model test.}\label{fig:ML9}
\end{center}
\end{figure}
\begin{figure}[t!]
\begin{center}
\includegraphics[width=8.0cm,angle=0]{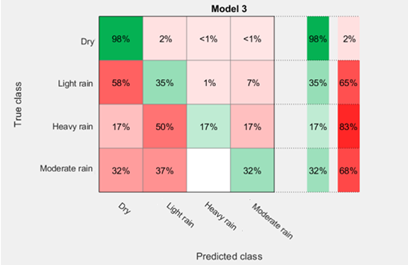}
\caption{Machine Learning results represented through the 
confusion matrix.}\label{fig:ML8}
\end{center}
\end{figure}

The last step of the training phase is the export of the model; subsequently, 
another data set was taken, for the testing phase of the Machine Learning 
algorithm, with a different time span from that of the training. The data were 
treated in the same way as the training data: in particular, they are processed
by using Excel, through the association of the physical quantities with the 
meteorological condition detected. Then, the data are imported on Matlab and 
finally statistical models for feature extraction (average, standard deviation 
and analysis of the main components) are applied. The input data for the 
model test are collectively shown in Fig.~\ref{fig:ML9}: in close analogy
with the training data, also in this case the hourly values of 
atmospheric pressure, maximum and
minimum temperature and humidity, and wind speed and direction are reported
over the whole temporal interval. 
The testing phase consists on the compilation of an algorithm for plotting the 
results. This algorithm is written on Matlab, and it is suited to make a 
comparison between the real conditions measured by the weather stations 
and the prediction that the trained model carries out. 
To evaluate the quality of an algorithm, it is possible to use tools making 
the Classification Learner available. A largely adopted example of these tools
is constituted by the confusion matrix reported in Fig.~\ref{fig:ML8}. 
The rows of this matrix indicate the model predictions, while columns refer to 
the data measured by the weather stations. Each matrix element show the 
correspondence (expressed in percentage) between predicted and observed 
data~\cite{Choi:16}.
The accuracy of the algorithm can be deduced by looking at the main diagonal,
which shows (in green) the correct predictions. All other (wrong) predictions,
corresponding to the other matrix elements, are reported in red.
In particular, in Fig.~\ref{fig:ML8} it can be 
seen that the dry condition is very well predicted, with an accuracy of 
98\%: this clearly depends on the amount of data chosen for testing the program,
where the dry days are the majority. This problem 
could be solved by extending the data acquisition time to a larger period 
and increasing the number of weather stations examined.
It is worth noting that, due to the 
approximation to unity, the sum of all the elements on a row does not exactly
match the 100\%. In addition, we also note that the element corresponding to
the fourth row and third column is white, since this particular combination
was never observed by the algorithm. 
\begin{figure}
\begin{center}
\begin{tabular}{cc}
\includegraphics[width=4.5cm,angle=0]{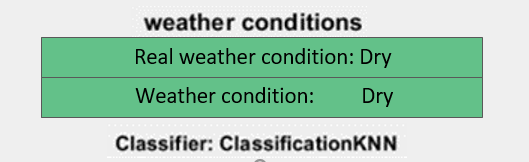} 
\includegraphics[width=4.5cm,angle=0]{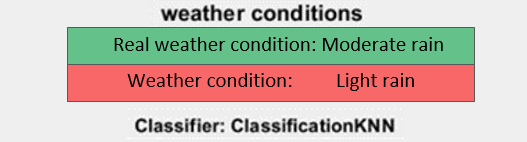}  \\
\includegraphics[width=4.5cm,angle=0]{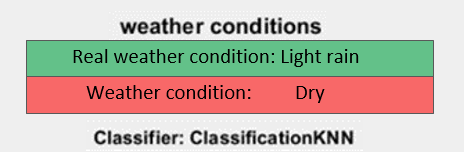} 
\includegraphics[width=4.5cm,angle=0]{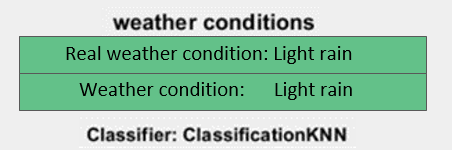}  \\
\end{tabular}
\caption{Accuracy of the Machine Learning predictions of the different
weather conditions.}\label{fig:ML10}
\end{center}
\end{figure}
The prediction performances can be summarized in a more intuitive way in
Fig.~\ref{fig:ML10}: here, when the correspondence between predicted and 
observed data is good, we can identify two green 
rectangles, which show real and the predicted weather conditions, 
respectively. According to the confusion matrix, 
the dry condition is almost always guessed. In the other panels of
Fig.~\ref{fig:ML10}
it is possible to note the existence of incorrect predictions, indicated 
by the red rectangles. Finally, in the last panel we have reported two
green rectangles also for the case of weak rain, since, after the dry case,
this is the best reprented condition, even though the accuracy is only
of 35\%.

\section{Conclusions}
In the present work we have presented a didactic approach suited to 
describe the Machine Learning application to the general problem of
weather forecast. 
In particular, we have been focused on the predictions of weather conditions
on geographic areas characterized by a complex orography, 
such the case of Sicily. A well known example is provided by the Etna
and Stromboli volcanoes, whose presence significantly influences 
the weather conditions, due to Stau and Foehn effects, with
possible impact on the air traffic of the nearby Catania and Reggio Calabria
airports.
We have shown that it is possible to use a simple thermodynamic
approach to calculate the final temperature of a mass of air undergoing an
adiabatic expansion or compression, such in the case of Stau and Foehn effects,
but no information are provided on the rainfall accumulation. For such
an aim we have proposed a Machine Learning approach which, while being only
at its initial formulation, is able to provide indication on the weather
conditions after a proper training phase with data input provided by
the Salina weather station. 
Specifically, in the case at issue we have shown that the
algorithm shows a great accuracy in predicting a dry condition, since the data
provided by the analyzed weather station registered mostly this particular
condition. The Machine Learing protocol described in the present work can 
be easily
improved, for instance by enriching it with further input data and  
enlarging the time span considered.

\section*{Acknowledgements}
The present work frames within the PON project titled ``Impiego di tecnologie, 
materiali e modelli innovativi in ambito aeronautico AEROMAT'', 
avviso1735/Ric, 13 luglio 2017. 

\bibliography{manuscript}
\bibliographystyle{ieeetr}
\end{document}